# A Study on Impact of Capital Structure on Profitability of Companies Listed in Indian Stock Exchange with respect to Automobile Industry

P. Aishwarya[1], *Sudharani R[2], *Dr. N Suresh[3]

[1]Student, M.Com in Accounting and Taxation
M S Ramaiah University of Applied Sciences
Karnataka, India
Tel: +91 9743801733

[2]Assistant Professor, Faculty of Management and Commerce
M S Ramaiah University of Applied Sciences
Karnataka, India
Tel: +91 8867658752

[3]Professor, Faculty of Management and Commerce
M S Ramaiah University of Applied Sciences
Karnataka, India
Email: [1]aishwaryapraveen78@gmail.com, [2] sudharani.co.mc@msruas.ac.in,
[3]nsuresh.ms.mc@msruas.ac.in

*Abstract:* *Current research helps in understanding both positive and negative impact of capital structure on profits of Indian automobile companies by using variables like Return on Capital Employed, Return on Long Term Funds, Return on Net Worth, Gross Profit Margin, Operating Profit and Return on Asset. The study hypothesized that RoCE, RoLT and RoNW has a positive effect and GP, OP and ROA has a negative impact on debt equity and interest coverage ratios i.e capital structure of the companies. Also the study proves that the relationship between profitability and capital structure variables are strongly significant. Hypothesis were tested by using fixed effect and random effect models by considering 10 years data (from 2010-2019) of 17 automobile companies. The result of the study recommends that the firms can improve their performance by using an optimal capital structure. Also a fair mix of debt and equity should be established to ensure that the firm maintains capital adequacy. Firms can thus be able to meet their financial compulsions and investments that can promise attractive returns.*

# 1. INTRODUCTION

Capital structure make best use of the market worth of a company that is if a company requiring an appropriately intended capital structures the collective worth of the rights and proprietorship benefits of the stockholders are exploited. Effective and efficient utilization of the capital structure bring about cost reduction. Appropriate blend of debt and equity enables the company to invest in profitable ventures. This is because capital structure upsurges the capability of the business to find new affluence by generating venture chances. With appropriate wealth gearing it also rises the self-confidence of dealers of debt. This enables firm to utilize leverage and enjoy the benefits of tax deduction, this leads to an increase in profitability. This is in line with a study conducted by (Friend and Lang) who established that there was a affirmative connection among capital structure and profitability. The findings revealed that firms that maintained an optimal capital structure obtained cheap funds to finance their operations which in turn generate returns and enhanced their financial performance.

Capital structure rises the nation's amount of venture and development by growing the company's chance to involve in forthcoming affluence-generating monies. This is because firms that make maximum use of leverage face attractive growth due increasing costs savings as a result of tax deduction. This is consistent with (Sarkar and Zapatero) who observed there was affirmative connection among leverage and productivity of businesses.

The capital structure of a company may be simple, compound or complex. A simple capital structure is composed of only one security base, for example, the equity share capital issued by a ,company. A compound capital structure indicates a combination of two security bases, in the form of equity and preference share capital. The complex capital structure is a mixture of multi-security bases, consisting of equity and performance share capital-and a series of debentures or bonds and loans from other sources. When a firm has a high level of business risk, it usually seeks to balance it with a lower level of financial risk by utilizing lower levels of debt capital in its capital structure. Determination of optimal capital structure is an important task in financial management. The term capital structure is different from financial structure. Financial structure refers to the way the firm's assets are financed. Capital structure is the permanent financing of the company. Thus, capital structure is part of financial structure.

In order to extend the understanding of the capital structure and its impact on the profitability of the company, employed the secondary data study methodology, sample of 17 automobile companies, spanning over a period of ten years. In doing so, we examine both; a) the impact capital structure with the firm performance and b) the effects of independent variable on the dependent variable**.**

This study makes two key contributions: first, it extends the understanding of capital structure and its factors and also the variables that effect the profitability of the company; second, it provides a more granular insight into the effects that industry actually face when the debt equity are taken into consideration.. In the next section we introduce the theoretical background and literature review. This is followed by the method employed in the research. Then present the results of the study and the hypothesis framing, followed by the discussion. Finally, we close the paper with conclusions and managerial implications

## 2. THEORETICAL BACKGROUND AND LITERATURE REVIEW

**2.1  Background Theory**

Capital structure choice is imperative for the firm, this is for the reason that it determines how well a firm can identify and invest in projects that can promise better returns. An investment decision made by the firm has an influence on its competitive abilities to cope with an aggressive environment (Wald). The capital structure of a company essentially is a blend of various securities. In broad, a company can go for amongst numerous options of capital structures. A firm can issue a huge quantity of debt. A firm can also organize to lease financing, use warrants, issue convertible bonds, sign forward contracts or trade bond swaps. Also it can issue dozens of different securities in limitless blends; nevertheless, it tries to get the exacting blend that makes best use of its general market worth (Hadlock and James).

**2.2  Critical review of literature should be based on:**

Various journal papers related to following topics are reviewed in this section
Impact of Capital Structure on Profitability of the company
The Relationship between Capital Structure & Profitability
Capital Structure and its Determinants of the Automobile Companies in India
The purpose of this study is to investigate the relationship between capital structure and profitability of ten listed Srilankan banks over the past 8 year period from 2002 to 2009.The data has been analysed by using descriptive statistics and correlation analysis to find out the association between the variables. Results of the analysis show that there is a negative association between capital structure and profitability except the association between debt to equity and return on equity. Further the results suggest that 89% of total assets in the banking sector of Sri Lanka are represented by debt, confirming the fact that banks are highly geared institutions. The outcomes of the study may guide banks, loan-creditors and policy planners to formulate better policy decisions as far as the capital structure is concerned (T. Velnampy & J. Aloy Niresh, 2012).

  The Capital Structure of a firm describes how it has sourced its finances. This capital structure is comprised of the owned & the owed capital. There are a number of determinants that affect the decisions taken while determining this capital structure like cost of capital, control, flexibility etc. The Indian Automobile Industry is the seventh-largest auto producer in the world with an average annual production of 17.5 Million vehicles.  This paper is an attempt to ascertain the impact of capital structure (CS) on the profitability (P) of the firm. Liquidity and growth in terms of performance of the firm have significant influence on debt-equity ratio. Capital structure, the mix of long term debts and equity securities, is generally used to finance long term assets of companies. It consists of permanent short-term debt, preferred stock, and common equity. The results revealed there is positive relationship between capital structure and financial performance (Dr. Atul A. Agwan, 2017).

  This paper talks about the Capital Structure of a firm describes how it has sourced its finances. This capital structure is comprised of the owned & the owed capital. There are varieties of determinants that have an effect on the choices taken whereas deciding this capital structure like cost of capital, control, flexibility etc. The Indian Automobile Industry is that the seventh-largest motor vehicle producer within the world with an average annual production of 17.5 Million vehicles. It's the fourth largest automotive market by volume, by 2015. It contributes concerning seven-membered to    the    country's value by    volume and comes 6 Million-plus vehicles to be sold-out annually, by 2020.In this paper, we have a tendency by aiming to study the capital structure of a number of the distinguished listed Indian Automobile firms. The target is to seek out the relationship between the capital structure, return on invested capital, value of the firm, and numerous different factors of the firm (Monik Shah, 2015).

This study examines the Capital structure and its determinants of Automobile companies listed in India using panel data analysis. The data was taken from secondary data source named as "Industry; financial aggregates and ratios" (PROWS) of centre for monitoring Indian economy (CMIE) covers 58 Indian Automobile companies listed on the Bombay Stock Exchange covering the period from 1997-98 to 2010-14 (17 years). Fixed effects regression model was used for the analysis of penal data of sample companies The empirical Results shows that the variables of profitability, size, tangibility, growth, and non-debt tax shield are negatively related with leverage and risk and liquidity are positively related with leverage. Profitability is statistically significant determinants of capital structure. While on the contrary, size, tangibility, growth, risk, non-debt tax shield and liquidity are statistically insignificant determinants of capital structure. The results are generally consistent with theoretical predictions as well as previous research papers. This paper adds to the existing literature on the relationship between the firm specific factors and leverage (Suresh Babu, 2016).

This study investigates the firm performance on capital structure for the listed non-financial companies in Dhaka Stock Exchange (DSE) for the period of 2008-2011 under judgment sampling method. Specific objective of this research is to examine the relationship between the attribute of capital structure and the performance as measured by Return on Assets (ROA) and Return on Sales (ROS). Multiple regression models were used to estimate the influence of capital structure on firm performance The results obtained from regression models show that Debt Ratio, Debt Equity Ratio and Proprietary of Equity Ratio are negatively and significant relationship with Return on Asset (ROA) and Return on Sales (ROS). Beside the control variable total asset is positively and significant relationship with Return on Asset (ROA) and Return on Sales (ROS) (Md. Abdur Rouf, 2015).

This study aims to find a relationship between the structure of capital & profitability. Various parameters namely Short-term & long-term Debts to Asset Ratio, Funded Capital Ratio, Funded Debt Ratio, Current Debt Ratio, Funded Asset Ratio & Sales Growth as an independent variable & Return on Assets of as dependent variable to find a relationship between Capital Structure & Profitability. 28 companies in Cement & Automobile sector of Pakistan Stock Exchange were chosen randomly as a sample. Secondary data for 7 years was collected from audited consolidated financial statements & analysed through descriptive statistical techniques namely Correlation & Regression. Housman test was used for selection of model. Results display both positive &negative relationship between the variables in Cement & Automobile sector (Adnan Ali, 2015).

## 3. PROBLEM STATEMENT

This chapter provides the research gap, title, aim and research objectives and scope employed to conduct the study. The chapter describes scope of the present research and methods and methodology used to collect the data and to analyse the same so as to meet the objectives of the study.

**3.1 Research Gap**

After examining several studies on the related topic, it was found that there was both positive and negative impact on the profitability of the company by examining the financial variables available in the financial statements. A review of prior studies reveals the absence of research on this topic. This paper will therefore fill the gap by taking up a comprehensive study of Indian Automobile Sector with specific factors such as Return on Asset, Return on Equity, Debt to Equity, Current Assets, Interest Coverage Ratio, Return on Net worth and Operating Profit Ratio to determine the continuous changes in regulatory framework of Automobile Industry and its impact on the profitability of those companies.

**3.2 Objectives**

The main objective of the study is to find the impact of capital structure on the profitability of the selected automobile Industries. The proposed research is intended to examine the trend and pattern of financing the capital structure of Indian automobile companies. Some specific objectives are as follows:

i. To study factors affecting the profitability of the company
ii. To identify and analyse the existing relationship between Capital Structure and Profitability
iii. To analyse the effect of Debt to Equity Ratio on Firms Performance
iv. To analyse the effect of Interest coverage Ratio on Firms Performance
v. To develop model and suggest suitable recommendations based on the outcomes

**3.3 Methodology**

The process of identifying the impact of capital structure on the profitability of the company started with searching of articles, journals. For the purpose of the study 17 automobile companies listed in Indian Stock Exchange were selected of which financial data were examined for getting the required data for the study. In order to achieve the above mentioned objectives the secondary data has been collected and implemented which are extracted from different journal, articles, books, published and unpublished sources and electronic databases and World Wide Web facilities. Published and unpublished documents of the organizations to be studied were also used for the purpose of knowing about what exactly Capital Structure means. The data were collected from money control website and certain company's websites for analysis. The present study is conducted to understand the impact on capital structure on the profitability of the automobile industry and also to study on the factors that affects the profitability of the firm. Then statistical tool is used for further analysis using Excel and E-Views 9.0 Student Version Software and inferences are drawn. Suitable graphs and tables are drawn to analyse descriptive data. Secondary data were used for the study. The required data were collected from moneycontrol website. The Automobile Industries i.e., Automobile- 2 & 3 Wheelers, Automobile- Auto &Truck Manufactures, Automobile- Dealers and Distributors, Automobile LCVs/HCVs, Automobile- Passenger Cars, Automobile- Tractors, Automobile- Trucks/LCVs which are listed in Indian Stock Exchange are selected as sampling design from 2010- 2019.

Table 1: List of companies selected for the purpose of the study

| REPRESENTING THE SAMPLE AUTOMOBILE COMPANIES LISTED IN INDIAN STOCK EXCHANGE | | |
|---|---|---|
| Sl.No. | Name of Automobile Company | Production |
| 1 | Bajaj Auto Limited | Automobile 2&3 Wheelers |
| 2 | Hero MotoCorp Company | Automobile 2&3 Wheelers |
| 3 | TVS Motor Company | Automobile 2&3 Wheelers |
| 4 | Atul Auto | Automobile 2&3 Wheelers |
| 5 | Scooters India | Automobile 2&3 Wheelers |
| 6 | Mahindra and Mahindra | **Automobile - Auto & Truck Manufacturers** |
| 7 | Tata Motors Limited | **Automobile - Auto & Truck Manufacturers** |
| 8 | Eicher Motors | **Automobile - LCVS/ HVCS** |
| 9 | Tata Motors LTD – DVR | **Automobile - LCVS/ HVCS** |
| 10 | Maruti Suzuki India | **Automobile - Passenger Cars** |
| 11 | Hindustan Motors | **Automobile - Passenger Cars** |
| 12 | Escorts | **Automobile - Tractors** |
| 13 | HMT | **Automobile - Tractors** |
| 14 | VST Tillers Tractors | **Automobile - Tractors** |
| 15 | Force Motors | **Automobile - Trucks/LCVs** |
| 16 | SML Isuzu | **Automobile - Trucks/LCVs** |
| 17 | Ashok Leyland | **Automobile - Trucks/LCVs** |

**Source**: Based on Authors' analysis

### 3.3.1 Panel Data Analysis

Panel data refers to a type of data that contains observations of multiple phenomena collected over different time period for the same group of individuals, units or entities. In short, in econometrics panel data refers to a multidimensional data collected over a period of time. Panel data analysis on the other hand refers to a statistical method widely used in different disciplines such as social sciences, econometrics to analyse data that are collected for multiple periods and over the same individuals or entities.

A simple panel data regression can specified thus:

$Y_{it} = a + bX_{it} + \varepsilon_{it}$

Y- Dependent variable

X-independent or explanatory variable

a, b- coefficients

i, t- indices for individuals and time

ε- Error term

Panel data regression analysis can be done in mainly three ways: (1) Independently Pooled OLS regression model. (2) Fixed effects model (3) Random effects model

**(1) Independently Pooled OLS regression model**

This type of panel data model assumes homogeneity of all sections of data in a panel data study that is it does not treat each section differently. Alternatively, it treats all section as just a single section of data. In short, there are no unique characteristics of individuals within the measurement set and no universal effects over time. Example if Panel data is collected for six countries across 1980-1990, Pooled OLS does not distinguish between these six countries while also neglecting the cross section and time series nature of the data.

**(2) Fixed Effects Model (FE)**

Fixed effects explore the relationship between predictor and outcome variables within an entity. Each entity has its own individual characteristics that may or may not influence the predictor variables. FE assumes that something within the individual may impact or bias the predictor or outcome variables and we need to control for this. FE removes the effect of those time in variant characteristics from the predictor variables so that we can assess the predictor's net effect. There is another important assumption of the FE that those time-invariant characteristics are unique to the individual units and should not be correlated with other individual characteristics. Each entity is different therefore the entity's error term and the constant should not be correlated with the others. According to Booth et al., (2001), the fixed effects model permits the use of all the available information, as the intercept can vary freely between firms and/or times. It can capture the effect of omitted variables through the change of intercept with the assumption that slope coefficients are constant across firms.

**(3)Random Effects Model (RE)**

Random effects model assumes that the individual or group effects are uncorrelated with other explanatory variables. It considers heterogeneity to be a variable that produces an impact on the regression residuals (Martin et al., 2005). If there are reasons to believe that differences across entities have some influence on the dependent variable then RE should be used.

**Hausman Test**

In order to choose between fixed and random effects models, Hausman test is applied. The underlined null hypothesis is that the preferred model is random effects against the fixed effects. It basically tests that whether or not the unique errors (ui) are correlated with the regressors. The null hypothesis assumes that is unique errors (ui) are not correlated with the regressors. If the test value is found to be significant, null hypothesis is rejected and use of fixed effects model is recommended.

# 4. RESULTS

**4.1 Data Analysis relating to Dependent Variable 1: Y1_DER**
**Hypothesis on testing appropriate model**
**Null Hypotheses:** Random Effect Model is appropriate
**Alternative Hypotheses:** Fixed Effect Model is appropriate
(Decision Criterion: Reject H0 if probability value is less than 5%, Accept H0 if probability value is greater than 5 %.)

**Table 2: Y1_DER Fixed Effects Model**

Dependent Variable: Y1_DER
Method: Panel Least Squares
Date: 06/02/20   Time: 09:22
Sample: 2010 2019
Periods included: 10
Cross-sections included: 17
Total panel (balanced) observations: 170

| Variable | Coefficient | Std. Error | t-Statistic | Prob. |
|---|---|---|---|---|
| C | 0.432532 | 0.076690 | 5.639984 | 0.0000 |
| GP | 7.57E-05 | 0.000133 | 0.567146 | 0.5715 |
| OP | -0.000221 | 0.000547 | -0.404501 | 0.6864 |
| ROA | 2.01E-05 | 0.000200 | 0.100123 | 0.9204 |
| ROCE | 0.004261 | 0.002066 | 2.063045 | 0.0409 |
| ROLT | -0.006964 | 0.002833 | -2.458485 | 0.0151 |
| RONW | -0.003232 | 0.001368 | -2.363228 | 0.0194 |

Effects Specification

Cross-section fixed (dummy variables)

| | | | |
|---|---|---|---|
| R-squared | 0.728111 | Mean dependent var | 0.327471 |
| Adjusted R-squared | 0.687420 | S.D. dependent var | 0.564162 |
| S.E. of regression | 0.315416 | Akaike info criterion | 0.655375 |
| Sum squared resid | 14.62464 | Schwarz criterion | 1.079630 |
| Log likelihood | -32.70686 | Hannan-Quinn criter. | 0.827533 |
| F-statistic | 17.89373 | Durbin-Watson stat | 0.893595 |
| Prob(F-statistic) | 0.000000 | | |

**Source**: Based on E-views output, Authors' analysis

**Table 3: Y1_DER Random Effects Model**

```
Dependent Variable: Y1_DER
Method: Panel EGLS (Cross-section random effects)
Date: 06/02/20   Time: 09:21
Sample: 2010 2019
Periods included: 10
Cross-sections included: 17
Total panel (balanced) observations: 170
Swamy and Arora estimator of component variances
```

| Variable | Coefficient | Std. Error | t-Statistic | Prob. |
|---|---|---|---|---|
| C    |  0.584440 | 0.068078 |  8.584807 | 0.0000 |
| GP   |  0.000192 | 0.000104 |  1.834579 | 0.0684 |
| OP   |  0.000586 | 0.000451 |  1.300771 | 0.1952 |
| ROA  | -0.000367 | 0.000163 | -2.256132 | 0.0254 |
| ROCE |  0.005281 | 0.001930 |  2.735583 | 0.0069 |
| ROLT | -0.012570 | 0.002041 | -6.158206 | 0.0000 |
| RONW | -0.004780 | 0.001252 | -3.817490 | 0.0002 |

**Effects Specification**

|  | S.D. | Rho |
|---|---|---|
| Cross-section random | 0.152380 | 0.1892 |
| Idiosyncratic random | 0.315416 | 0.8108 |

**Weighted Statistics**

| | | | |
|---|---|---|---|
| R-squared | 0.300969 | Mean dependent var | 0.179347 |
| Adjusted R-squared | 0.275238 | S.D. dependent var | 0.402488 |
| S.E. of regression | 0.342650 | Sum squared resid | 19.13763 |
| F-statistic | 11.69666 | Durbin-Watson stat | 0.680950 |
| Prob(F-statistic) | 0.000000 | | |

**Unweighted Statistics**

| | | | |
|---|---|---|---|
| R-squared | 0.496817 | Mean dependent var | 0.327471 |
| Sum squared resid | 27.06576 | Durbin-Watson stat | 0.481485 |

**Source**: Based on E-views output, Authors' analysis

Out of the above two effects (Fixed and Radom effect), only one effect is appropriate. To check the appropriateness Correlated Random Effects- Hausman Test has been done.

**Table 4: Y1_DER Hausman Test**

```
Correlated Random Effects - Hausman Test
Equation: Untitled
Test cross-section random effects
```

| Test Summary | Chi-Sq. Statistic | Chi-Sq. d.f. | Prob. |
|---|---|---|---|
| Cross-section random | 35.362409 | 6 | 0.0000 |

Cross-section random effects test comparisons:

| Variable | Fixed | Random | Var(Diff.) | Prob. |
|---|---|---|---|---|
| GP   |  0.000076 |  0.000192 | 0.000000 | 0.1629 |
| OP   | -0.000221 |  0.000586 | 0.000000 | 0.0092 |
| ROA  |  0.000020 | -0.000367 | 0.000000 | 0.0009 |
| ROCE |  0.004261 |  0.005281 | 0.000001 | 0.1656 |
| ROLT | -0.006964 | -0.012570 | 0.000004 | 0.0043 |
| RONW | -0.003232 | -0.004780 | 0.000000 | 0.0049 |

```
Cross-section random effects test equation:
Dependent Variable: Y1_DER
Method: Panel Least Squares
Date: 06/02/20   Time: 09:01
Sample: 2010 2019
Periods included: 10
Cross-sections included: 17
Total panel (balanced) observations: 170
```

| Variable | Coefficient | Std. Error | t-Statistic | Prob. |
|---|---|---|---|---|
| C    |  0.432532 | 0.076690 |  5.639984 | 0.0000 |
| GP   |  7.57E-05 | 0.000133 |  0.567146 | 0.5715 |
| OP   | -0.000221 | 0.000547 | -0.404501 | 0.6864 |
| ROA  |  2.01E-05 | 0.000200 |  0.100123 | 0.9204 |
| ROCE |  0.004261 | 0.002066 |  2.063045 | 0.0409 |
| ROLT | -0.006964 | 0.002833 | -2.458485 | 0.0151 |
| RONW | -0.003232 | 0.001368 | -2.363228 | 0.0194 |

**Effects Specification**

Cross-section fixed (dummy variables)

| | | | |
|---|---|---|---|
| R-squared | 0.728111 | Mean dependent var | 0.327471 |
| Adjusted R-squared | 0.687420 | S.D. dependent var | 0.564162 |
| S.E. of regression | 0.315416 | Akaike info criterion | 0.655375 |
| Sum squared resid | 14.62464 | Schwarz criterion | 1.079630 |
| Log likelihood | -32.70686 | Hannan-Quinn criter. | 0.827533 |
| F-statistic | 17.89373 | Durbin-Watson stat | 0.893595 |
| Prob(F-statistic) | 0.000000 | | |

**Source**: Based on E-views output, Authors' analysis

In this case we can see that the null hypotheses i.e., random effect model is appropriate is rejected and the alternative hypotheses i.e., fixed effect model is appropriate is been accepted as the probability value are less than 5% and is significant with the dependent variable.

**Observation:** From the above table profitability has a significant relationship along with the positive trend with y1_der (Debt Equity Ratio) of Automobile industry. This R-squared measures the strength of the relationship between the model and the dependent variable on a convenient scale. Here in this case the R-squared value is 0.728 i.e., 72.8% which means the model is fitting good and reveals that there is 72.8% variation in debt equity ratio was as result of profitability of automobile industry over the period of study. Overall, there is a positive effect towards the capital structure of the Automobile industry.

**4.2: Wald Test**

Wald test is a way to find out if explanatory variables in a model are significant or not or they have long term effect or short term effect between the variables.

**Null hypothesis: There is no short-run relationship among the variables**
**Alternative hypothesis: There is a short-run relationship among the variables**

**Table 5: Y1_DER Wald test 1**

Wald Test:
Equation: Untitled

| Test Statistic | Value | df | Probability |
|---|---|---|---|
| t-statistic | 0.567146 | 147 | 0.5715 |
| F-statistic | 0.321654 | (1, 147) | 0.5715 |
| Chi-square | 0.321654 | 1 | 0.5706 |

Null Hypothesis: C(2)=0
Null Hypothesis Summary:

| Normalized Restriction (= 0) | Value | Std. Err. |
|---|---|---|
| C(2) | 7.57E-05 | 0.000133 |

Restrictions are linear in coefficients.

**Table 6: Y1_DER Wald test 2**

Wald Test:
Equation: Untitled

| Test Statistic | Value | df | Probability |
|---|---|---|---|
| t-statistic | 0.467763 | 150 | 0.6406 |
| F-statistic | 0.218802 | (1, 150) | 0.6406 |
| Chi-square | 0.218802 | 1 | 0.6400 |

Null Hypothesis: C(3)=0
Null Hypothesis Summary:

| Normalized Restriction (= 0) | Value | Std. Err. |
|---|---|---|
| C(3) | 1.317179 | 2.815911 |

Restrictions are linear in coefficients.

**Source**: Based on E-views output, Authors' analysis

In the table shown above we can see that probability value is more than 0.05%, which shows that null hypothesis is not rejected as there is a long term effect where Operating profit and Return on Assets cannot jointly influence the dependent variable Debt equity ratio.

**4.3: Testing of Hypotheses**
**Hypothesis on testing the relationships of variables**

**Null hypothesis:** "There is no significant relationship between debt equity ratio and gp, op, roa, roce, rolt and ronw"
**Alternative Hypothesis 1:** "There is a significant relationship between Debt Equity Ratio and Gross Profit"
**Alternatives Hypothesis 2:** "There is a significant relationship between Debt Equity Ratio and Operating Profit"
**Alternative Hypothesis 3:** "There is a significant relationship between Debt Equity Ratio and Return on Asset"
**Alternatives Hypothesis 4:** "There is a significant relationship between Debt Equity Ratio and Return on Capital Employed"
**Alternative Hypothesis 5:** "There is a significant relationship between Debt Equity Ratio and Return on Long Term Funds"
**Alternatives Hypothesis 6:** "There is a significant relationship between Debt Equity Ratio and Return on Net worth"

**4.4: Data Analysis relating to Dependent Variable 2: Y2_ICR**
**Hypothesis on testing appropriate model**

Null Hypotheses: Random Effect Model is appropriate
Alternative Hypotheses: Fixed Effect Model is appropriate
(Decision Criterion: Reject H0 if probability value is less than 5%, Accept H0 if probability value is greater than 5 %.)

**Table 7: Y2_ICR Fixed Effects Model**

Dependent Variable: Y2_ICR
Method: Panel Least Squares
Date: 06/02/20   Time: 09:30
Sample: 2010 2019
Periods included: 10
Cross-sections included: 17
Total panel (balanced) observations: 170

| Variable | Coefficient | Std. Error | t-Statistic | Prob. |
|---|---|---|---|---|
| C | 212.1063 | 85.60335 | 2.477780 | 0.0144 |
| GP | 0.010397 | 0.148964 | 0.069797 | 0.9444 |
| OP | 0.575614 | 0.610648 | 0.942629 | 0.3474 |
| ROA | 0.018763 | 0.223645 | 0.083895 | 0.9333 |
| ROCE | -0.403826 | 2.305678 | -0.175144 | 0.8612 |
| ROLT | 1.519161 | 3.161772 | 0.480478 | 0.6316 |
| RONW | 0.224590 | 1.526490 | 0.147128 | 0.8832 |

Effects Specification

Cross-section fixed (dummy variables)

| | | | |
|---|---|---|---|
| R-squared | 0.832295 | Mean dependent var | 283.5027 |
| Adjusted R-squared | 0.807196 | S.D. dependent var | 801.8187 |
| S.E. of regression | 352.0742 | Akaike info criterion | 14.69078 |
| Sum squared resid | 18221567 | Schwarz criterion | 15.11504 |
| Log likelihood | -1225.717 | Hannan-Quinn criter. | 14.86294 |
| F-statistic | 33.16081 | Durbin-Watson stat | 2.609567 |
| Prob(F-statistic) | 0.000000 | | |

**Source**: Based on E-views output, Authors' analysis

**Table 8: Y2_ICR Random Effects Model**

Dependent Variable: Y2_ICR
Method: Panel EGLS (Cross-section random effects)
Date: 06/12/20   Time: 16:19
Sample: 2010 2019
Periods included: 10
Cross-sections included: 17
Total panel (balanced) observations: 170
Swamy and Arora estimator of component variances

| Variable | Coefficient | Std. Error | t-Statistic | Prob. |
|---|---|---|---|---|
| C | 182.8005 | 209.7728 | 0.871422 | 0.3848 |
| GP | 0.014887 | 0.145278 | 0.102474 | 0.9185 |
| OP | 0.636188 | 0.600289 | 1.059803 | 0.2908 |
| ROA | 0.013688 | 0.219829 | 0.062264 | 0.9504 |
| ROCE | -0.249695 | 2.291138 | -0.108983 | 0.9134 |
| ROLT | 2.631676 | 3.034814 | 0.867162 | 0.3871 |
| RONW | 0.091787 | 1.516071 | 0.060543 | 0.9518 |

Effects Specification

| | | S.D. | Rho |
|---|---|---|---|
| Cross-section random | | 794.8013 | 0.8360 |
| Idiosyncratic random | | 352.0742 | 0.1640 |

Weighted Statistics

| | | | |
|---|---|---|---|
| R-squared | 0.540004 | Mean dependent var | 39.32902 |
| Adjusted R-squared | 0.036206 | S.D. dependent var | 354.6502 |
| S.E. of regression | 348.1708 | Sum squared resid | 19759330 |
| F-statistic | 2.058113 | Durbin-Watson stat | 2.398491 |
| Prob(F-statistic) | 0.060910 | | |

Unweighted Statistics

| | | | |
|---|---|---|---|
| R-squared | 0.108448 | Mean dependent var | 283.5027 |
| Sum squared resid | 96869240 | Durbin-Watson stat | 0.489243 |

**Source**: Based on E-views output, Authors' analysis

Out of the above two effects (Fixed and Radom effect), only one effect is appropriate. To check the appropriateness Correlated Random Effects- Hausman Test has been done.

**Table 9: Y2_ICR Hausman Test**

```
Correlated Random Effects - Hausman Test
Equation: Untitled
Test cross-section random effects

Test Summary                    Chi-Sq. Statistic   Chi-Sq. d.f.   Prob.
Cross-section random                 2.405694            6          0.8789

Cross-section random effects test comparisons:

Variable        Fixed         Random      Var(Diff.)     Prob.
GP            0.010397      0.014887      0.001085      0.8916
OP            0.575614      0.636188      0.012544      0.5886
ROA           0.018763      0.013688      0.001692      0.9018
ROCE         -0.403826     -0.249695      0.066835      0.5510
ROLT          1.519161      2.631676      0.786709      0.2097
RONW          0.224590      0.091787      0.031700      0.4557

Cross-section random effects test equation:
Dependent Variable: Y2_ICR
Method: Panel Least Squares
Date: 06/02/20   Time: 09:33
Sample: 2010 2019
Periods included: 10
Cross-sections included: 17
Total panel (balanced) observations: 170

Variable     Coefficient   Std. Error    t-Statistic    Prob.
C            212.1063      85.60335      2.477780      0.0144
GP           0.010397      0.148964      0.069797      0.9444
OP           0.575614      0.610648      0.942629      0.3474
ROA          0.018763      0.223645      0.083895      0.9333
ROCE        -0.403826      2.305678     -0.175144      0.8612
ROLT         1.519161      3.161772      0.480478      0.6316
RONW         0.224590      1.526490      0.147128      0.8832

                      Effects Specification
Cross-section fixed (dummy variables)

R-squared              0.832295    Mean dependent var      283.5027
Adjusted R-squared     0.807196    S.D. dependent var      801.8187
S.E. of regression     352.0742    Akaike info criterion   14.69078
Sum squared resid      18221567    Schwarz criterion       15.11504
Log likelihood        -1225.717    Hannan-Quinn criter.    14.86294
F-statistic            33.16081    Durbin-Watson stat      2.609567
Prob(F-statistic)      0.000000
```

**Source**: Based on E-views output, Authors' analysis

In this case we can see that the null hypotheses i.e., random effect model is appropriate is been accepted and the alternate hypotheses i.e., fixed effect model is appropriate been rejected. We came to know that Random Effect Model is appropriate by doing Hausman test were the probability value is 0.878 i.e., 87.8% which is above 5% so we do not reject the null hypotheses rather we accept it.

**Observation:** From the above table profitability does not have significant relationship with y2_icr (Interest Coverage Ratio) of Automobile industry. This R-squared measures the strength of the relationship between the model and the dependent variable on a convenient scale. Here in this case the R-squared value is 0.540 i.e., 54% which means the r-squared is less but the overall model is fitting and has variation in interest coverage ratio was as result of profitability of automobile industry over the period of study.

**4.5: Testing of Hypotheses**
**Hypothesis on testing the relationships of variables**

**Null hypothesis:** There is no significant relationship between interest coverage ratio and gp, op, roa, roce, rolt and ronw
**Alternative Hypothesis 1:** "There is a significant relationship between Interest Coverage Ratio and Gross Profit"
**Alternatives Hypothesis 2:** "There is a significant relationship between Interest Coverage Ratio and Operating Profit"
**Alternative Hypothesis 3:** "There is a significant relationship between Interest Coverage Ratio and Return on Asset"
**Alternatives Hypothesis 4:** "There is a significant relationship between Interest Coverage Ratio and Return on Capital Employed"
**Alternative Hypothesis 5:** "There is a significant relationship between Interest Coverage Ratio and Return on Long Term Funds"
**Alternatives Hypothesis 6:** "There is a significant relationship between Interest Coverage Ratio and Return on Net worth"

# 5. DISCUSSION AND CONCLUSION

This research is an extension of previous research, where a set of capital structure variables is considered to examine their association with the firm performance. From the above research work the study concludes, on comparison with existing literature we can see that most of the results of the present study are giving the results which support some literatures. On the other hand, a few findings of the current study also contradict some of the Contentions of previous studies. The objective of this study was to identify the capital structure that affects the firm performance and examine the relationship between the attribute of capital structure and the performance of listed automobiles companies in Indian Stock Exchange. These capital structure include Debt Equity Ratio (DER), Interest Coverage Ratio (ICR) of the firm and profitability include Gross Profit (GP), Operating Profit (OP), Return on Asset (RoA), Return on Capital Employed (RoCE), Return on Long Term Funds (RoLT) and Return on Net worth (RoNW).

This study used panel data of 17 automobile companies for the period of 10 years that is from 2010-2019 creating 170 observations of the data. Researcher analysed the impact of capital structure variables (dependent variables) against profitability variables (independent variable). Pooled OLS Regression model was used to analyse the overall result of the dependent and the independent variable after which Fixed effect model and Random effects model after which Hausman specification test was done to check the best model suitable for the variables selected for the purpose of the study. For dependent variable i.e., debt equity fixed effect model was suitable and Wald test was done for the three variables to check whether it has long term effect or short term effect in nature. And for dependent variable i.e., interest coverage random effect model was suitable as it was above 5% was used to measure the relationship between capital structure and profitability of the automobile industries. Moreover, partial correlation technique also used to measure the relationship between the study variables in order to support the regression results. After testing the relationship between the variables, researcher revealed the mixed results between capital structure variables and company profitability that means capital structure variables indicated a negative relationship with company profitability variables at certain stages while other capital structure variables indicated a positive relationship with profitability variables at certain point of time. Debt equity ratio and Interest coverage ratios were used as capital structure indicators of automobile industries. The Fixed effect model result indicated a significant and positive relationship with some of the independent variables namely RoCE, RoLT and RoNW whereas it had a long term effect on the variables namely GP, OP and RoA which was explained by doing Wald test. In Fixed effect model case the R-squared value is 0.728 i.e., 72.8% which means the model is fitting good and reveals that there is 72.8% variation in debt equity ratio was as result of profitability and the p-value is 0.00 which is less than 5% which suggests that there is a significant relationship between the variables of capital structure and the profitability of automobile industry over the period of study. Overall, there is a positive effect towards the capital structure of the Automobile industry.

The Random effect model result indicated that there is no significance and positive relationship with some of the independent variables namely GP, ROA, RoCE, and RoNW whereas OP and RoLT had a significant relationship between the variables of capital structure. In Random effect model case the R-squared value is 0.070 i.e., 7 % which means the r-squared is very less but the overall model is fitting good and has variation in interest coverage ratio was as result of profitability of automobile industry over the period of study.

The random effect model results indicated a negative relationship between return on capital employed (RoCE) against interest coverage ratio (ICR) at a coefficient of -0.40 which was also supported by the correlation results at -0.30 that means there is no relationship between capital structure and company profitability in terms of return on capital employed. But in the case of fixed effect model results indicated a positive relationship between return on capital employed and the debt equity ratio at 0.004 coefficient level which was also supported by the correlation results at 0.08.

Finally this study revealed that, capital structure of listed automobile companies in Indian Stock Exchange affect company profitability in terms of the variables positively**.** On the other side, capital structure of listed automobile companies has negative relationship with company profit in terms of return on capital employed. The results indicate that debt usage has more advantage for companies that depend much on assets to generate profit than those that depend much on equity or shareholders fund to generate company profit.

**5.1: Suggestions**

The present study aimed at making a critical study of the impact of capital structure on profitability of the Indian Automobile Industries. After analysing the various aspects of the capital structure and the profitability of selected company few suggestions are made:

From the above study it can be suggested that automobile companies need not necessarily concentrate on the debt-equity ratio to maximise its profitability.

From the study it was seen that the return on capital employed is one of the ratio that has gone negative in case of interest coverage ratio. Thus, it can be suggested that the companies should concentrate on their operational efficiency to improve its interest coverage ratio.

It is suggested that the financial managers should look into the future prospects of the companies while designing the capital structure policies, as it affects the price of its shares for a longer period as seen in the case of Operating profit and the return on asset.

**5.2: Limitations of the study**

The study is limited to only 17 Companies for a period of 10 years. Therefore, this comprises the result of only a few numbers of variables of automobile firms, which would not be sufficient to totally generalize the inference of the automobile industry. The data used for the study are secondary in nature. Therefore, the accuracy of the results of analysis is totally dependent upon the reliability and accuracy of secondary data. Only secondary data are collected to analysis to do this research.

The limitation of the study is used only automobile companies as a sample. So, the results may not extend across all listed companies in stock Exchange. The study explores only six firm performance or profitability variables, other factors influencing the capital structure of the firm such as Return On Equity (ROE), Return On Investment (ROI) and Tobin's Q ratio could be explored in further studies.

Here the company's financial performance or profitability is computed based on debt equity, interest coverage, return on asset, return on capital employed, return on net worth, return on long term funds but too many factors or measures have impact on financial performance of companies. So the result will be further valuable when researcher considers varies other kinds of parameters. The sample size used in this research is relatively small considering the total number of listed companies in National Stock exchange.